\documentclass[aip, jcp, preprint]{revtex4-2}

\usepackage{amsfonts,amsmath,amssymb,bm}
\usepackage{graphicx}
\usepackage[english]{babel}
\usepackage{color}

\usepackage{url}

\usepackage{orcidlink}

\begin{document}

\title{Artificial-intelligence-based surrogate solution of dissipative quantum dynamics: 
physics-informed reconstruction of the universal propagator}

\author{Jiaji Zhang\,\orcidlink{0000-0003-2978-274X}}
\email{jiaji.zhang@zhejianglab.com}
\affiliation{Zhejiang Laboratory, Hangzhou 311100, China}

\author{Carlos L. Benavides-Riveros\, \orcidlink{0000-0001-6924-727X}}
\email{cl.benavidesriveros@unitn.it}
\affiliation{Pitaevskii BEC Center, CNR-INO and Dipartimento di Fisica, Università di Trento, I-38123 Trento, Italy }

\author{Lipeng Chen\, \orcidlink{0000-0003-1415-4767}}
\email{chenlp@zhejianglab.com}
\affiliation{Zhejiang Laboratory, Hangzhou 311100, China}

\begin{abstract}
The accurate (or even approximate) solution of the equations that govern the dynamics of dissipative quantum systems remains a challenging task for quantum science. While several algorithms have been designed to solve those equations with different degrees of flexibility, they rely mainly on highly expensive iterative schemes. Most recently, deep neural networks have been used for quantum dynamics but current architectures are highly dependent on the physics of the particular system and usually limited to population dynamics. Here we introduce an artificial-intelligence-based surrogate model that solves dissipative quantum dynamics by parameterizing quantum propagators as Fourier neural operators, which we train using both dataset and phy\-sics-informed loss functions. Compared with conventional algorithms, our quantum neural propagator avoids time-consuming iterations and provides a universal super-operator that can be used to evolve any initial quantum state for arbitrarily long times. To illustrate the wide applicability of the approach, we employ our quantum neural propagator to compute population dynamics and time-correlation functions of the Fen\-na-Matthews-Olson complex.
\end{abstract}

\maketitle

\section{Introduction}
\label{sec.introduction}

Since atomic and molecular systems are never completely isolated from the surrounding environment, a proper description of their dynamics requires the theory of open quantum systems \cite{breuer2007, weiss2012}. The environment is usually treated as a heat bath, and thus a commonly used strategy in the theory of open quantum systems is to trace out bath degrees of freedom and derive the corresponding equations of motion (EOM) for the system, also referred to as quantum master equations (QMEs) \cite{devega2017,Alicki2003}. Under perturbative and Markovian approximations, the best-known QMEs are the Lindblad \cite{lindblad1976,Gorini76} and Redfield \cite{redfield1957} equations. Numerically exact QME approaches, such as the hierarchy equations of motion \cite{ishizaki2005ltc,tanimura2020} and the quasiadiabatic propagator path integral \cite{QUAPI1,QUAPI2}, are developed to capture the intrinsic non-perturbative and non-Markovian system dynamics. The modern technique of ultrafast spectroscopy provides essential insights into the intricate system-bath interactions and bath-induced electronic dephasing and energy relaxation processes \cite{mukamel1999,mukamel2009, gelin2022review}. The key quantity for simulating optical spectroscopy is polarization, which is linked to various response functions made calculable by QMEs \cite{10.1063/1.5006824}. QMEs approaches have also been utilized to simulate both linear and nonlinear spectra of molecular systems \cite{tanimura2009, may2011,ishizaki2012}.          


Apart from their different concrete forms or range of validity, a common ingredient of the EOM for open quantum systems is the presence of a partial differential equation with respect to time which is usually well-suited for iterative numerical solvers \cite{butcher2016}. 
The most widely used numerical solvers are the family of Runge-Kutta methods, which iteratively integrate the EOM with a small enough time step $(\delta t)$ from a given initial condition.
As a universal formalism, no restrictions are a priori made on the explicit form of the EOM or the initial states. Runge-Kutta methods thus provide a versatile tool to solve EOM for a wide variety of scenarios such as rate equations in chemical reactions \cite{ishizaki2009redfield,zhang2020pt}, population dynamics in charge and energy transfer processes \cite{may2011,  ishizaki2009heom}, and spectroscopic simulations \cite{zhang2021pcet, ikeda2017qhfp}. 
Despite its straightforward im\-ple\-men\-tation, the drawbacks of Runge-Kutta methods are the large computational cost and the presence of numerical instabilities associated with the iterations, especially for long-time dynamics.
Improvements have been proposed to overcome these limitations by deriving modified schemes of EOM {\cite{nazir2009, kimura2014, mulvihill2022qme,PhysRevLett.127.270503,Wang2023}} or implementing other numerical solvers (e.g., proper orthogonal decomposition {\cite{liu2023dmd}}). However, since they are usually limited to a few specific scenarios, an efficient universal solver is still to be proposed.

Over recent years, artificial intelligence (AI) has emerged as a powerful tool that, inte\-gra\-ted into research, is augmenting and accelerating scientific discovery \cite{LeCun2015}. It is also increasingly being seen as a useful tool to solve various physical and chemical problems that have remained so far very challenging for conventional methods \cite{hermann2023,natchem2020}.
Deep neural networks are the best-known example of AI tools applied to quantum systems. The universal approximation theorem of (Borel-measurable) functions {\cite{hornik1989}} provides the basis for the neural-network parameterization of wavefunctions for quantum spin systems {\cite{sharir2020, carleo2017rbm}}. 
When further extended to many-electron systems, they outperformed traditional variational ans\"atze {\cite{schatzle2023, pfau2020, cassella2023, hermann2023}}. Yet, while they are somehow routinely applied to study static (ground-state) properties, there are relatively few attempts to simulate the dynamics of quantum systems \cite{PhysRevLett.125.100503,PhysRevLett.127.230501,Ullah2022,PhysRevLett.128.090501}, and existing dynamical deep neural-network models are usually highly dependent on the physics of specific systems, limited to population dynamics \cite{Ullah2022}, or tested on toy models only \cite{PhysRevLett.122.250502,PhysRevLett.122.250503,PhysRevLett.122.250501}.

While the aforementioned works are concerned with the approximation of quantum states, AI can also be used as a surrogate model to approximate highly nonlinear operators. Based on the universal approximation theorem of functionals and mathematical operators {\cite{chen1995}}, the solution operator of a given partial differential equation can be parameterized as a deep neural network that describes the mapping between initial states and the evolved state at some subsequent time {\cite{kovachki2023fno, lu2022don, rosofsky2023}.
Recently developed architectures, such as DeepONet {\cite{lu2021don, lu2022don}} and Fourier Neural Operator (FNO) {\cite{li2021fno, kovachki2021fno}}, have demonstrated their superiority on classical differential equations in weather forecasting {\cite{jaideep2022fcn, jiang2021dte}} and latent diffusion models in text-to-image transformation {\cite{guibas2021afno}}. Despite this progress, surrogate models for quantum operators have only been applied, to the best of our knowledge, to scattering processes  \cite{PhysRevD.108.L101701}.

In this work, our objective is to develop an AI-based surrogate model for the universal solution of EOM for open quantum systems. To this end, we train a neural operator as the (universal) quantum propagator of dissipative quantum dynamics using both dataset and physics-informed loss functions. We employ the FNO architecture to parameterize the propagator for the Lindblad QME, which is the quantum analog of the solution operator in a typical surrogate model. Our approach is quite versatile and can be easily extended to solve other EOM for quantum dynamics. The trained propagator allows for the direct computation of dynamics up to a chosen time limit $t_{\mathrm{max}}$ for any initial state through a single-step operation without invoking the tedious, expensive iterations.

Moreover, since the neural propagator obeys the usual composition property of quantum operators, the method can easily be extended to arbitrarily long times. In addition to the simulation of population dynamics, our neural propagator can be used to compute the more challenging time-correlation functions of system operators.
We test our approach by training a neural operator as the quantum propagator of the well-known Fen\-na-Matthews-Olson complex and compute population dynamics as well as multi-time correlation functions of system operators.

\section{Result}

\subsection{The QME propagator and time-correlation functions}

We consider the typical Fen\-na-Matthews-Olson pigment-protein complex found in green sulfur bacteria as our model system \cite{ adolphs2006fmo, duan2022fmo}.
The interplay between molecular excitations, environment interaction, and quantum coherence effects makes the complex one of the most important workhorses for developing and improving simulation methods \cite{Kim2020,Milder2010,Engel2007}.
The dissipative dynamics of the reduced density operator $\hat{\rho}$ for the electronic sub-system is described by the Lindblad QME,
\begin{equation}
\frac{\partial}{\partial t} \hat{\rho}(t) = -\frac{i}{\hbar} \left[\hat{H}_{el}, ~ \hat{\rho}(t) \right] -
\sum_{j=1}^{N} \frac{\lambda_{j}}{2} \left( \hat{V}_{j}^{\dagger} \hat{V}_{j}\hat{\rho}(t) +  \hat{\rho} (t)\hat{V}_{j}^{\dagger} \hat{V}_{j} 
-2\hat{V}_{j} \hat{\rho}(t) \hat{V}_{j}^{\dagger} \right) ,
\label{eq.qme}
\end{equation}
where $\hat{H}_{el}$ is the Hamiltonian of the electronic states {\cite{ishizaki2009heom}}
\begin{equation}
\hat{H}_{el} = \sum_{j=1}^{N} \varepsilon_{j} |{j}\rangle \langle{j}| +
\sum_{j \ne j^{\prime}} \Delta_{j, j^{\prime}} |{j}\rangle \langle{j^{\prime}}|,
\end{equation}
with $\varepsilon_{j}$ being the energy of the $j$-th state $|j\rangle$ (there are $N$ of them) and $\Delta_{j,j^{\prime}}$ the inter-state couplings. The second part on the right-hand side of Eq.~\ref{eq.qme} represents a pure-dephasing Lindblad operator with $\hat{V}_{j} = |j\rangle \langle j|$ and a pure-dephasing rate $\lambda_j$. 

To alleviate the notation, we introduce the abbreviated index $x = (j, j^{\prime})$ and align the matrix entries of the reduced density operator as column vector, $\vec{\rho}_{t} = \left\{ \rho(x_0, t), \rho(x_1, t), ... \right\}$, with $\rho(x,t) = \langle{j}| \hat{\rho}(t) |{j^{\prime}}\rangle$.
Eq.~\eqref{eq.qme} can be recast to a matrix-vector form as
\begin{equation}
\frac{\partial}{\partial t} {\vec{\rho}}_t  = {\bm{L}} {\vec{\rho}}_t ,
\label{eq.qme.mv}
\end{equation} 
where the matrix entries of ${\bm{L}}$ is inferred from the right-hand side of Eq.~\eqref{eq.qme}.

The QME propagator ${\bm{G}}_t$ is defined through the integration of Eq.~{\eqref{eq.qme.mv}} as 
\begin{equation}
{\vec{\rho}}_t = {\bm{G}}_t {\vec{\rho}}_0 = e^{t {\bm{L}}} {\vec{\rho}}_0,
\label{eq.prop.def}
\end{equation}
where ${\bm{G}}_t$ can be regarded as a time-dependent matrix that always acts on the vector to its right.
The evaluation of dynamics up to arbitrarily long times can be inferred from the well-known composition property of quantum operators as
\begin{equation}
{\vec{\rho}}_{t_1+t_2} = {\bm{G}}_{t_2} {\vec{\rho}}_{t_1} = {\bm{G}}_{t_2} {\bm{G}}_{t_1} {\vec{\rho}}_{0}.
\label{eq.prop.comp}
\end{equation}
The first-order and second-order time-correlation functions (TCFs) are defined through the propagator as (see ``Methods'' for more details):
\begin{equation}
R^{(1)}(t_1) = {\bm{X}}_{\rm tr} {\bm{G}}_{t_1} {\bm{X}}_{\times} {\vec{\rho}}_{0},
\label{eq.resp.1st}
\end{equation}
\begin{equation}
R^{(2)}(t_1, t_2) = {\bm{X}}_{\rm tr} {\bm{G}}_{t_2} {\bm{X}}_{\times} {\bm{G}}_{t_1}{\bm{X}}_{\times}{\vec{\rho}}_{0},
\label{eq.resp.2nd}
\end{equation}
where  ${\bm{X}}_{\times}$ and ${\bm{X}}_{\rm tr}$ recover the operations for any operator $\hat{X}$ as
\begin{equation}
{\bm{X}}_{\times} \vec{\rho} = \frac{i}{\hbar} \left[ \hat{X}, \,  \hat{\rho} \right], 
\end{equation}
\begin{equation}
{\bm{X}}_{\rm tr} \vec{\rho} = \sum_{j} \langle j| \hat{X}  \hat{\rho} | j \rangle .
\end{equation}
Conventional methods such as Runge-Kutta evaluate Eqs.~{\eqref{eq.resp.1st}} and {\eqref{eq.resp.2nd}} by starting from an initial state ${\vec{\rho}}_{0}$ and iteratively propagate Eq.~{\eqref{eq.qme.mv}} for all combinations of $t_{1}$ and $t_{2}$. The overall computational cost thus scales exponentially with the order of TCFs, limiting, as a consequence, iteration-based methods to low-order cases.

\subsection{Neural propagator within the FNO architecture}

Within the AI surrogate framework, we use the FNO architecture to parametrize the propagator ${\bm{G}}_t(\theta)$ ($\theta$ denote the set of trainable parameters) up to a chosen time limit $t_{\mathrm{max}}$. 
The model takes the initial density matrix ${\vec{\rho}}_{0}$ and time $t \in [0, t_{\mathrm{max}}]$ as the input, and outputs the density matrix ${\vec{\rho}}_t(\theta) = {\bm{G}}_t(\theta) \vec{\rho}_{0}$ that satisfies Eq.~{\eqref{eq.qme.mv}}.
As this parametrization retains the composition property of Eq.~{\eqref{eq.prop.comp}}, the model can be extended far beyond $t_{\mathrm{max}}$ by iteratively applying the neural operator.
In addition, the constructed FNO model works for any form of ${\vec{\rho}}_{0}$, including for instance the  ${\bm{X}}_{\times} {\vec{\rho}}_{0}$.
Thus, to make use of the full power of the AI-based architecture we will substitute the algebraic propagator in Eqs.~{\eqref{eq.resp.1st}} and {\eqref{eq.resp.2nd}} by the neural propagator ${\bm{G}}_t(\theta)$.

To train the FNO model, we prepare a dataset by first randomly sampling a set of initial states and then evaluating their time evolution using fourth-order Runge-Kutta (RK4) with a chosen time step $\delta t$. 
The dataset loss function is defined as follows:
\begin{equation}
\mathcal{L}_{\mathrm{data}} = \int_0^{t_{\mathrm{max}}} dt\, \frac{\left|\left|  {\bm{G}}_t(\theta) \vec{\rho}_{0} -  \vec{\rho}_{t} \right|\right|_{F}}
{\left|\left| {\bm{G}}_t(\theta) \vec{\rho}_{0} \right|\right|_{F}},
\label{eq.data_loss}
\end{equation}
where $\vec{\rho}_{t}$ is the corresponding data, and $||\cdot||_{F}$ denotes the Frobenius-norm.
In addition, we add to this loss function a physics-informed part, defined as
\begin{equation}
\mathcal{L}_{\mathrm{phys}} = \int_0^{t_{\mathrm{max}}} dt\frac{ \left|\left|  \frac{\partial}{\partial t}{\bm{G}}_t(\theta) \vec{\rho}_{0}- 
\bm{L} ({\bm{G}}_t(\theta) \vec{\rho}_{0} )\right|\right|_{F}}{\left|\left|\bm{L} ({\bm{G}}_t(\theta) \vec{\rho}_{0}) \right|\right|_{F}} 
+ \frac{ \left|\left| {\bm{G}}_{t=0}(\theta) \vec{\rho}_{0} - \vec{\rho}_{0} \right|\right|_{F}}{\left|\left| \vec{\rho}_{0} \right|\right|_{F}}.
\label{eq.phys_loss}
\end{equation}
The first term is the residual of Eq.~{\eqref{eq.qme.mv}}, and the second term is introduced to ensure the model reproduces the identity when $t=0$.
Importantly, notice that only the inputs $\vec{\rho}_{0}$ are required for the evaluation of the losses in $\mathcal{L}_{\mathrm{phys}}$. We can thus employ the on-the-fly sampling to generate those additional initial states.
More details on the model and the training can be found in Section ``Methods''.

\subsection{Applications to dissipative quantum dynamics}

We apply the trained FNO propagator to the excitation energy transfer of the seven-site Fen\-na-Matthews-Olson complex. 
To test our model, we compare results for population dynamics and TCFs with those from the RK4. 
Fig.~{\ref{fig.population}} shows the population dynamics obtained from both FNO propagator and RK4 up to  $50 t_{\mathrm{max}}$ with $t_{\mathrm{max}} = 30$fs.
Here we consider two typical initial states: 
an excitation located on (a) site 1, i.e., $\hat{\rho}_0=|{1}\rangle \langle{1}|$ 
and (b) site 6, i.e., $\hat{\rho}_0=|{6}\rangle \langle{6}|$.
Following the experimentally demonstrated energy transfer pathway {\cite{ishizaki2009heom, duan2022fmo}}, we focus on the time evolution of populations $p_{n}(t) = \langle n | \hat{\rho}_t|n\rangle$ for sites 
(a) $n =$ 1, 2, and 3, and (b) $n = $ 4, 5, and 6. 
Within the range $t \in [0, 50 t_{\mathrm{max}}]$, the FNO propagator yields population dynamics in excellent agreement with those obtained from the RK4. In both cases, the results are essentially exact up to $20 t_{\mathrm{max}}$. It is thus quite remarkable that our FNO propagator can infer truly long-time dynamics from the relatively short-time training data we used (recall that the propagator was trained with data in the interval  $[0, t_{\mathrm{max}}]$).

\begin{figure}[t]
\centering
\includegraphics[width=0.9\textwidth]{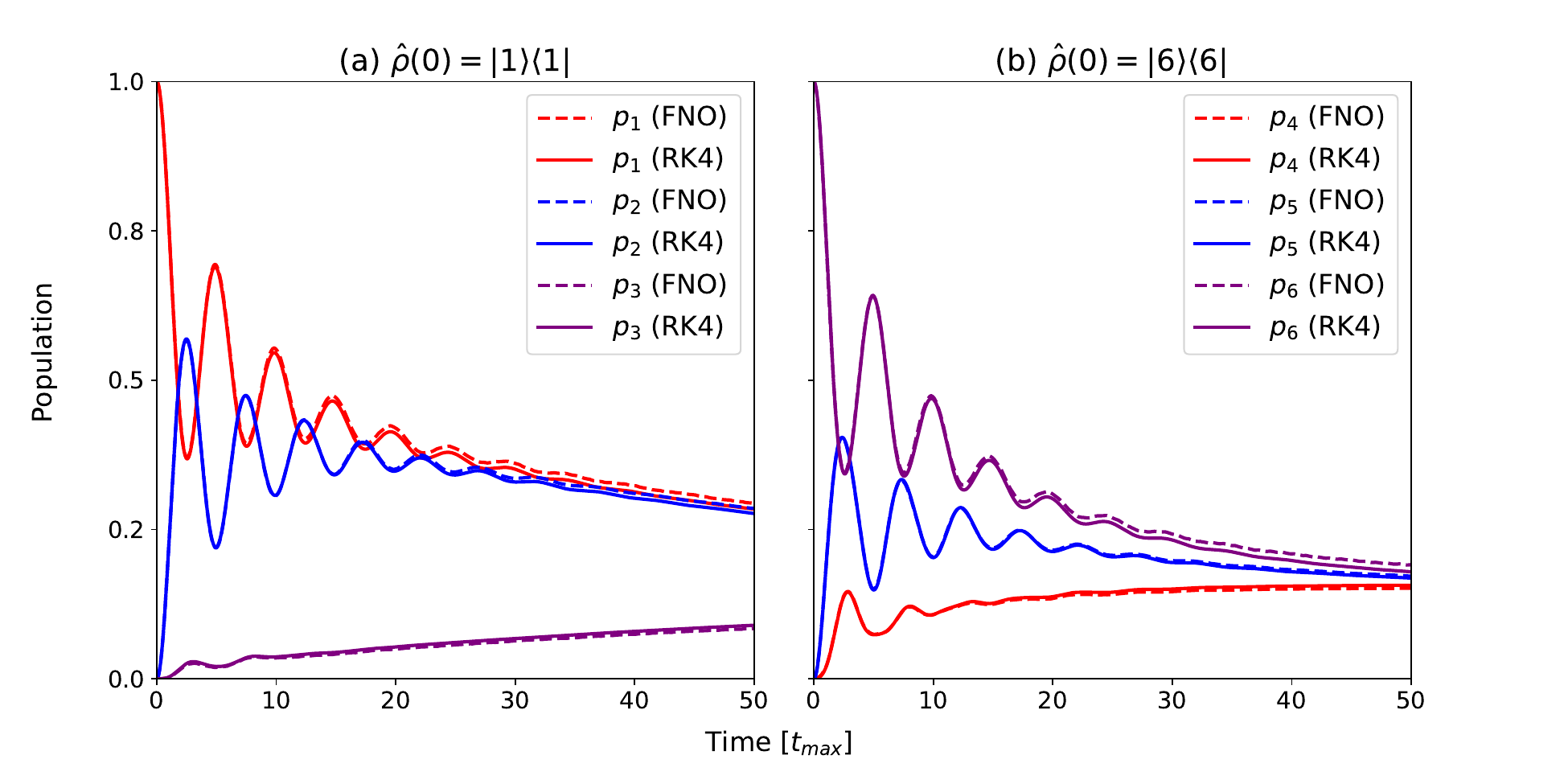}
\caption{Population dynamics of the FMO complex computed using RK4 (solid line) and FNO neural propagator (dashed line) for two initial states.
The time is presented in units of the maximum training time $t_{\mathrm{max}} = 30$fs.}
\label{fig.population}
\end{figure}

Next, we apply the FNO propagator to calculate first-order and second-order TCFs  using the initial condition of $\hat{\rho}_0=|1\rangle\langle{1}|$ and the system operator
\begin{equation}
\hat{X} = \sum_{j=1}^{N-1} \left( |{j}\rangle\langle{j+1}| + |{j+1}\rangle\langle{j}| \right),
\end{equation}
which is arbitrarily chosen for the demonstration of our algorithm. Although we can choose any other type of operator (e.g., to measure the correlation of populations of different sites or even coherences), the rich structure of $\hat{X}$, which corresponds to the hopping term of the lattice Hamiltonian, makes it an excellent test-bed for evaluating the performance of our neural propagator when tasked with computing TCFs. To facilitate the comparison of numerical results, we focus on the Fourier transform of TCFs over the time variables.

\begin{figure}
\centering
\includegraphics[width=12cm]{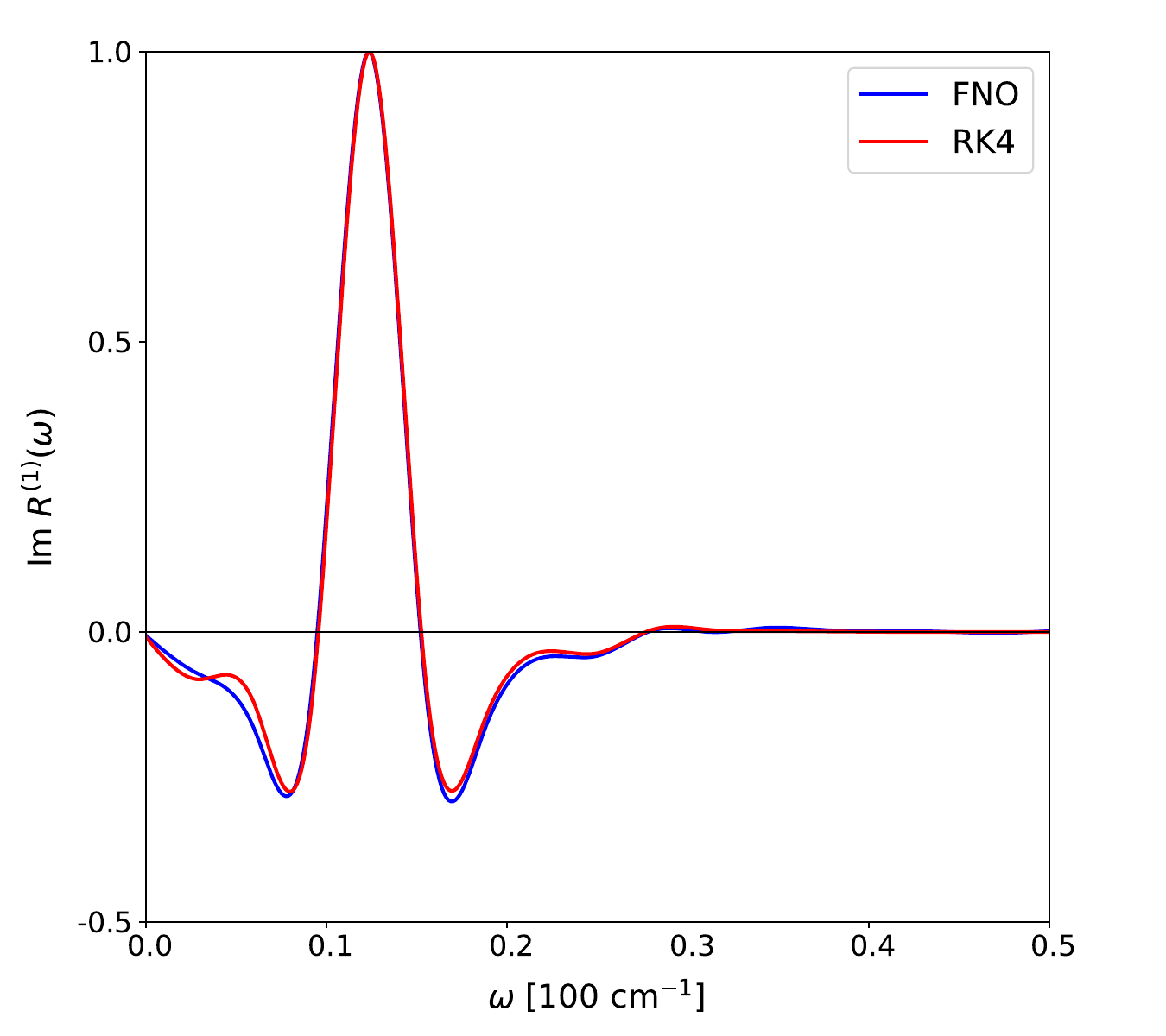}
\caption{The imaginary part of Fourier transform of the first-order TCF \eqref{eq.resp.1st}. The blue and red curves are results obtained from the FNO propagator and the RK4, respectively. In each case, the peak intensities are normalized with respect to their maximum value.}
\label{fig.linear_response}
\end{figure}

In Fig.~\ref{fig.linear_response}, we show the imaginary part of Fourier transform of the first-order TCF ($t_1 \in [0, 50 t_{\mathrm{max}}]$) evaluated using both the FNO propagator and the RK4. For each case, we normalize peak intensities with respect to their maximum value. 
In Fourier domain, the spectrum obtained from the FNO propagator matches perfectly with that from the RK4. 
The small deviation around 5 $\rm{cm^{-1}}$ may be attributed to the specific architecture of the FNO. It is important to note that the quantum system in this work is described by matrix entries over a discrete Hilbert space, which may differ in nature from the classical system in continuous coordinate space. Indeed, it has been pointed out that in certain circumstances, applying the FNO architecture (which relies on Fourier transformations) to discontinuous systems may yield non-optimal solutions {\cite{lu2022don}}. Nevertheless, overall, our FNO propagator produces results in excellent agreement with those obtained using the RK4 method.

In Fig.~{\ref{fig.second_response}}, we show the imaginary part of Fourier transform of the second-order TCF ($t_1,t_2 \in [0, 40 t_{\mathrm{max}}]$). 
As defined in Eq.~{\eqref{eq.resp.2nd}}, the presence of the second operator $\bm{X}_{\times}$ between propagators $\bm{G}_{t_1}(\theta)$ and $\bm{G}_{t_2}(\theta)$ generates additional transitions during propagation.
Accurate simulation thus requires the propagator to be universally applicable to any density matrix.
As shown in Fig.~{\ref{fig.second_response}}(a) and {\ref{fig.second_response}}(b), the results obtained from the FNO propagator are again in good agreement with those from the RK4, featuring very precise peak localization. 
Despite the complexity of the signal, the mean absolute error 
(i.e., $| R^{(2)}_{\rm{FNO}}-R_{\rm{RK4}}^{(2)}|$) is just $4.8\times 10^{-3}$. Still, we identify a slight tendency of the neural propagator to overestimate the envelopes of the peaks.   
It's worth noting that computing the second-order TCF takes only a few minutes when using the FNO propagator, whereas, with the RK4 method, it requires around one hour due to the iterations involved. 
The extension to the calculations of TCFs involving more time variables is straightforward, and one can reasonably expect an even more prominent computational speed-up. In our opinion, this demonstrates the broad applicability of our FNO propagator.

\begin{figure}
\centering
\includegraphics[width=\textwidth]{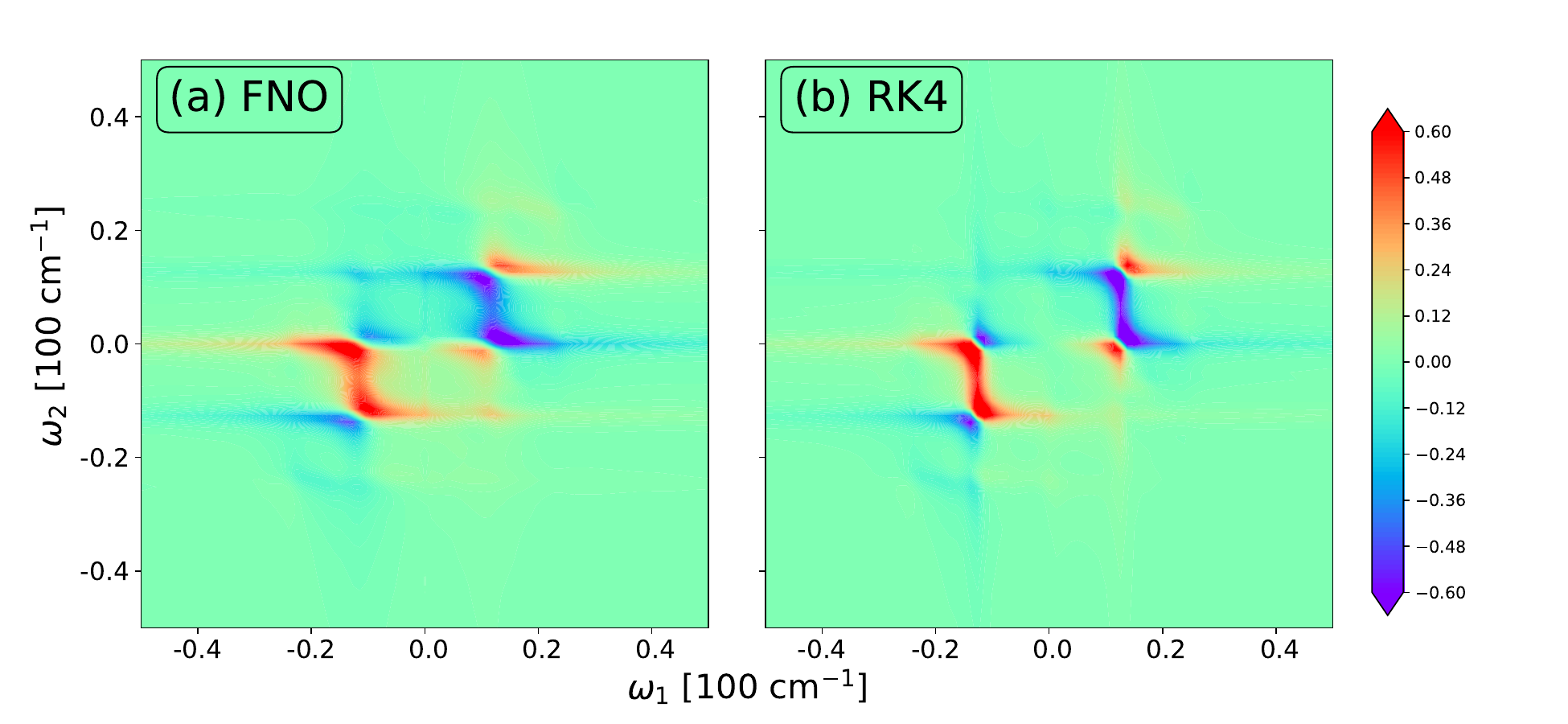}
\caption{The imaginary part of Fourier transform of the second-order TCF \eqref{eq.resp.2nd}. 
The panels (a) and (b) are results obtained from the FNO propagator and the RK4.}
\label{fig.second_response}
\end{figure}


\section{Discussion and outlook}
\label{sec.conclusion}

In this work, we develop an AI-based surrogate model to simulate dissipative quantum dynamics. Unlike typical deep neural networks that parametrize quantum states, our surrogate model parameterizes quantum propagators as universal neural operators, which are capable of mapping initial quantum states to their evolved counterparts at any subsequent time. The neural operator is trained using both dataset and physics-informed loss functions that ensure the universality of the trained propagator. The accuracy of the surrogate model is gauged by calculating both population dynamics and multi-time correlation functions of the Fenna-Matthews-Olson complex, yielding results in excellent agreement with those from conventional RK4. Quite remarkably, our FNO propagator, besides avoiding the conventional time-consuming iterations, captures quantum dynamics far beyond the small time window chosen in our training sets. 

This work lays the foundation for further exploration of neural propagators. Currently, our focus is on the time-independent case, where the dynamics depends solely on the Hamiltonian. Future investigations may extend to scenarios involving driven quantum systems, where the time evolution is influenced by the interaction with the external fields or by non-Markovian dynamics 
\cite{Alicki2003,10.1063/5.0051101}. In addition, in many non-dissipative scenarios, using wave functions alone may suffice. For such physical settings, we expect an excellent performance of our neural propagators (which were applied here to non-pure quantum states). Furthermore, our AI-based approach could be extended to compute photoemission spectra or transport dynamics in ab-initio quantum chemistry, where many existing numerical methods such as the $GW$ approximation, coupled-cluster techniques, or nonequilibrium Green’s functions, still face several shortcomings at both the fundamental and practical levels \cite{PhysRevLett.131.216401,PhysRevLett.129.066401,PhysRevLett.130.246301,PhysRevB.95.155203,YIN2022101843}.

\section{Methods}

\subsection{The model Hamiltonian and propagator}

The Fen\-na-Matthews-Olson protein complex is modeled as an electronic system interacting with heat baths which consist of a set of harmonic oscillators. The total Hamiltonian is defined as 
\begin{equation}
\hat{H}_{tot} = \hat{H}_{el} + \hat{H}_{ph}+ \hat{H}_{el-ph}  
+ \hat{H}_{reorg}.
\label{eq.Hamiltonian}
\end{equation} 
The first term $\hat{H}_{el}$ is the Hamiltonian of the electronic system for Fenna-Matthews-Olson complex (in ${\rm{cm^{-1}}}$) \cite{adolphs2006fmo}:
\begin{equation}
\hat{H}_{el} = 
\begin{bmatrix}
12410 & -87.7 & 5.5 & -5.9 & 6.7 & -13.7 & -9.9 \\
-87.7 & 12530 & 30.8 & 8.2 & 0.7 & 11.8 & 4.3 \\
5.5 & 30.8 & 12210 & -53.5 & -2.2 & -9.6 & 6.0 \\
-5.9 & 8.2 & -53.5 & 12320 & -70.7 & -17.0 & -63.6 \\
6.7 & 0.7 & -2.2 & -70.7 & 12480 & 81.1 & -1.3 \\
-13.7 & 11.8 & -9.6 & -17.0 & 81.1 & 12630 & 39.7 \\
-9.9 & 4.3 & 6.0 & -63.3 & -1.3 & 39.7 & 12440
\end{bmatrix}.
\end{equation}
The second term is the Hamiltonian of phonon heat baths,
\begin{equation}
\hat{H}_{ph} = \sum_{j=1}^{N} \sum_{\alpha} \left( \frac{\hat{p}_{j,\alpha}^2}{2} 
+ \frac{\omega_{j,\alpha}^2 \hat{x}_{j,\alpha}^2}{2} \right) ,
\end{equation}
where $\hat{p}_{j,\alpha}$, $\hat{x}_{j,\alpha}$, and $\omega_{j,\alpha}$ are the dimensionless momentum, coordinate, and frequency of the $\alpha$-th phonon mode of the $j$-th bath. The third term is the exciton-phonon coupling Hamiltonian,
\begin{equation}
\hat{H}_{el-ph} = -\sum_{j=1}^{N} \hat{V}_{j} \sum_{\alpha} c_{j, \alpha} \hat{x}_{j,\alpha},
\end{equation} 
where $c_{j,\alpha}$ represents the coupling constant between the $j$-th state and the $\alpha$-th phonon mode. 
Finally, the last term of Eq.~\eqref{eq.Hamiltonian} serves as the counter term,
\begin{equation}
\hat{H}_{reorg} = \sum_{j=1}^{N} \lambda_{j} |{j}\rangle\langle{j}| , 
\end{equation} 
with $\lambda_{j}$ being the reorganization energy of the $j$-th electronic state.
Under the perturbative and Markovian approximation for the heat baths, the dissipative dynamics is described by the Lindblad QME, as defined in Eq.~{\eqref{eq.qme}}. 
Throughout this work, we choose $\lambda_{j} = 35 {\rm{cm^{-1}}}$, following the experimentally determined value at high temperature {\cite{ishizaki2009heom}}.

The first-order and second-order TCFs for any  operator $\hat{X}$ are defined as
\begin{equation}
R^{(1)}(t_1) = \frac{i}{\hbar} \left\langle \left[\bar{X}(t_1), \, \bar{X}(0) \right] \right\rangle,
\end{equation}
\begin{equation}
R^{(2)}(t_1, t_2) = {\left( \frac{i}{\hbar}\right)}^{2} \left\langle \left[\left[\bar{X}(t_2), \, \bar{X}(t_1) \right], \, \bar{X}(0) \right] \right\rangle,
\end{equation}
where $[\hat{A}, \, \hat{B}] = \hat{A}\hat{B} - \hat{B}\hat{A}$, 
$\bar{X}(t) = e^{i\hat{H}_{tot} t/\hbar} \hat{X} e^{-i\hat{H}_{tot} t/\hbar}$ is the Heisenberg representation of operator $\hat{X}$, and $\langle \cdot \rangle$ denotes the ensemble average over the density matrix of total system.
Using the cyclic invariance, the above equations can be recast as \cite{mukamel1999}  
\begin{equation}
\label{R1}
R^{(1)}(t_1) = \frac{i}{\hbar} {\rm{Tr}} \left\{ \hat{X} \hat{G}_{tot}(t_1) \hat{X}^{\times} \hat{\rho}_{tot}(0) \right\}, 
\end{equation}
\begin{equation}
\label{R2}
R^{(2)}(t_1, t_2) = {\left( \frac{i}{\hbar} \right) }^{2}{\rm{Tr}} \left\{ \hat{X} \hat{G}_{tot}(t_2) \hat{X}^{\times} \hat{G}_{tot}(t_1) \hat{X}^{\times} \hat{\rho}_{tot}(0) \right\},
\end{equation}
where $\hat{\rho}_{tot}(0)$ and $\hat{G}_{tot}(t) = \exp( - i \hat{H}_{tot}^{\times} t /\hbar)$ are the initial density operator and the quantum propagator of the total system, and we have introduced the notation $\hat{A}^{\times} \hat{B} = \hat{A} \hat{B} - \hat{B} \hat{A}$. 
To obtain a reduced description of Eqs.~\ref{R1} and \ref{R2}, one traces the bath degrees of freedom, leading to the formalism with $\hat{\rho}_{tot}(0)$ and $\hat{G}_{tot}(t)$ replaced by the reduced density operator $\hat{\rho}_{0}$ and the reduced propagator $\hat{G}(t)$, respectively (see Eq.~{\eqref{eq.qme}}). 
The first-order TCF can be evaluated as follows. Starting from $\hat{\rho}_0$, we update the matrix entries by $\hat{X}^{\times}\hat{\rho}_0 \to \hat{\rho}^{\prime}_0$, and propagate $\hat{\rho}^{\prime}_0$ up to $t_{1}$ using iterative method (Eq.~{\eqref{eq.qme}}). $R^{(1)}(t_1)$ is then obtained by computing the trace of $\hat{X}\hat{\rho}^{\prime}_{t_1}$ for a series of $t_{1}$. To obtain the second-order TCF, we update $\hat{\rho}^{\prime}_{t_1}$ by $\hat{X}^{\times}\hat{\rho}^{\prime}_{t_1}\to \hat{\rho}^{''}_{t_1}$, propagate up to $t_{2}$ and compute the trace of $\hat{X}\hat{\rho}^{''}_{t_1, t_2}$. The overall computational cost thus increases dramatically with the number of time variables, which restricts the iteration-based propagation methods (such as the RK4) to lower-order cases.

\subsection{Quantum propagator in the FNO architecture}

\begin{figure}
\centering
\includegraphics[width=\textwidth]{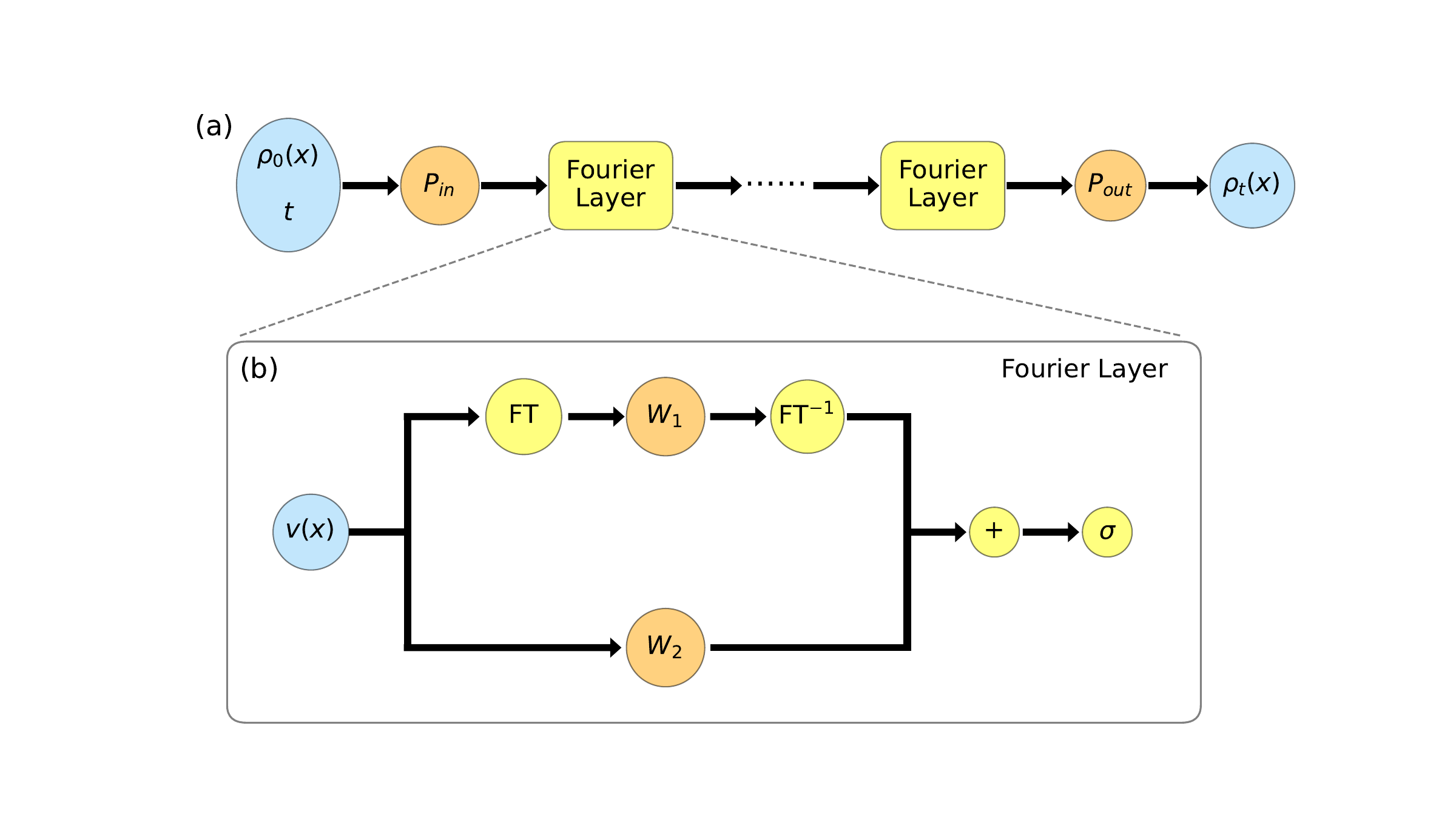}
\caption{The architecture of the FNO propagator taken from Ref.~{\cite{li2021fno}}. $P_{in}$ and $P_{out}$ are linear projection layers. $W_{1}$ and $W_{2}$ are two different convolution layers. $\mathrm{FT}$ and $\mathrm{FT}^{-1}$ denote the Fourier transform and its inverse. $+$ and $\sigma$ represent the element-wise sum and the activation function. The learnable parameters are those in $P_{in}$, $P_{out}$, $W_{1}$, and $W_{2}$.}
\label{fig.fno_model}
\end{figure} 

The vanilla FNO architecture is used in this work to parametrize the propagator ${\bm{G}}_t(\theta)$ with a set of trainable parameters $\theta$ for time $t$ up to a chosen maximum time limit $t_{\mathrm{max}}$. The model architecture is shown in Fig.~{\ref{fig.fno_model}}. 
In Fig.~{\ref{fig.fno_model}} (a), $P_{in}$ and $P_{out}$ denote the linear projections between the physical and latent spaces, which are parametrized as linear multilayer perceptrons with one hidden layer and a rectified linear unit (ReLU) activation function. The structure of the Fourier layer is shown in Fig.~{\ref{fig.fno_model}} (b), where $\mathrm{FT}$, $\mathrm{FT}^{-1}$, $+$, and $\sigma$ denote the Fourier transform and its inverse, element-wise sum, and ReLU, respectively. On the upper route, the linear convolution $W_{1}$ is applied to the layer-wise input in Fourier space, while on the lower route, another convolution $W_{2}$ is directly applied without FT.
For $W_{1}$, only the lowest $k_{\mathrm{max}}$ Fourier modes are explicitly included in the convolution, while all the modes with higher frequencies are truncated for numerical stability.
The results from two routes are summed and activated by ReLU before passing to the next layer. The learnable parameters $\theta$ are those in $P_{in}$, $P_{out}$, $W_{1}$, and $W_{2}$. Note that all the parameters $\theta$ are complex-valued, and ReLU is separately applied to the real and imaginary parts as $\sigma(a+ib) = \sigma(a) + i \sigma(b)$.

The FNO model takes the initial density matrix $\vec{\rho}_{t_0}$ and time $t_1$ as input and outputs $\vec{\rho}_{t_0 + t_1}$ corresponding to the solution of the QME. The dynamics can be easily extended to arbitrarily long times by using the composition property of quantum operators. 
Moreover, no restrictions are a priori made on the explicit form of the initial states $\vec{\rho}_{t_0}$. We can thus directly employ the FNO propagator to calculate TCFs by taking ${\bm{X}}_{\times} {\vec{\rho}}_{t_0}$ as input and inferring the dynamics up to $t_1$ and $t_{2}$. 
Each inference requires running a forward pass of the model, which needs far less computational cost than conventional iterative methods such as the RK4.

\subsection{Training and validation tests}
We use 8 layers to construct the neural propagator and parameterize the projection parts $P_{in}$ and $P_{out}$ as 2-layer feedforward neural networks with a hidden channel of size 512. 
The others are all Fourier layers, each of which has 32 Fourier modes and a hidden channel of size 256. The dataset was prepared by randomly sampling initial density matrices, which are drawn from the Gaussian Unitary Ensemble. 
We prepare in total 400 random samples, 200 of which are used as the training set, and the rest are held back as the validation set. The EOM of Eq.~\ref{eq.qme} is integrated by the RK4 for time up to $t_{\mathrm{max}} = 30 {\rm{fs}}$ with a time step of $\delta t = 0.6 {\rm{fs}}$. We train the neural propagator with up to $2000$ epochs using Adam optimizer and loss functions defined in Eqs.~\eqref{eq.data_loss} and \eqref{eq.phys_loss}.
All the computation is carried out on a single Nvidia 4090 GPU with 24GB memory.

\begin{figure}
\centering
\includegraphics[width=0.75\textwidth]{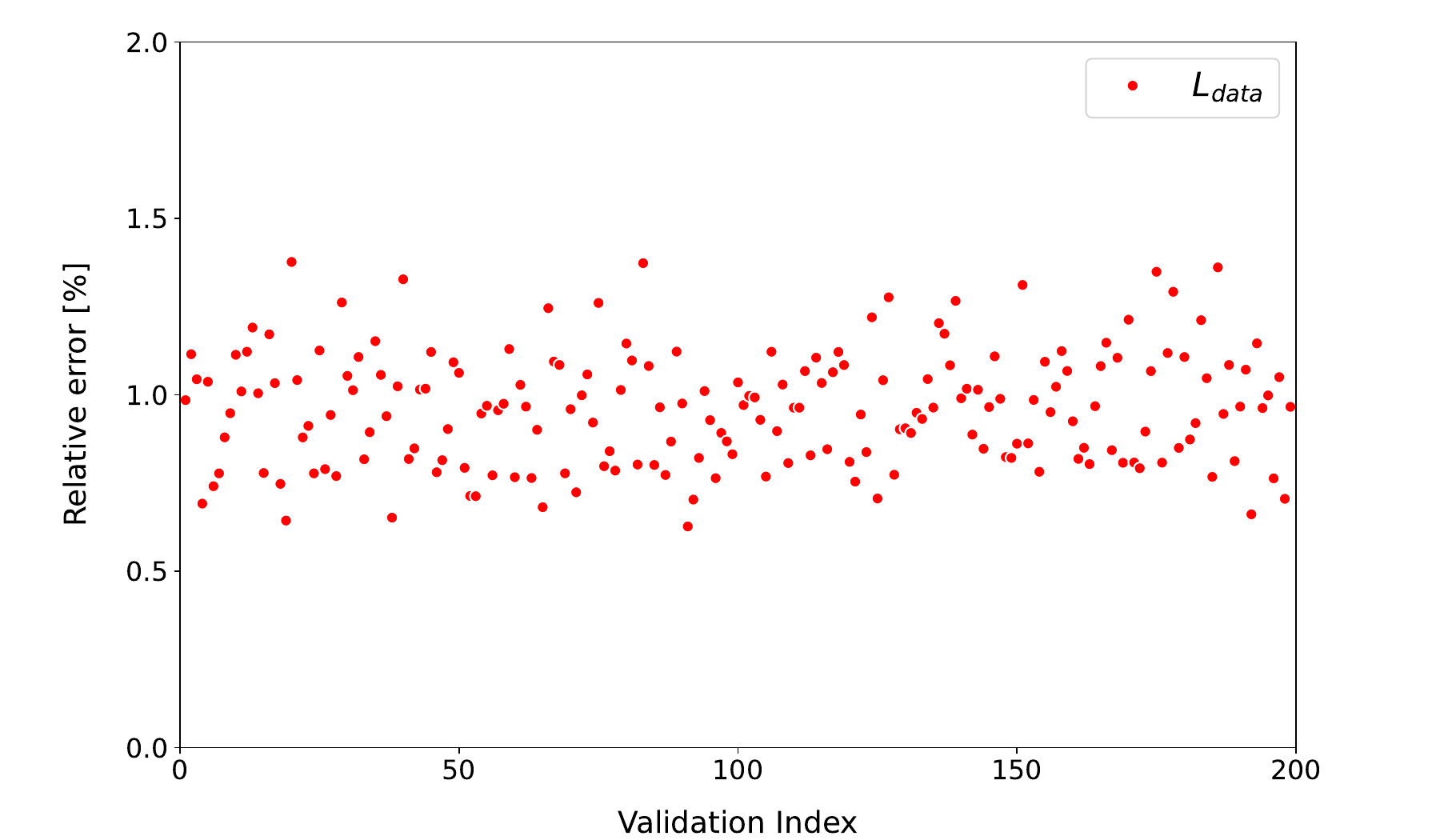}
\caption{The relative error of the data loss $\mathcal{L}_{\mathrm{data}}$ for the validation set. The horizontal axis represents the index of the samples contained in the validation set.}
\label{fig.validation}
\end{figure}

We now briefly discuss the on-the-fly sampling algorithm for $\mathcal{L}_{\mathrm{phys}}$.
We first prepare an initial dataset by randomly sampling an additional 400 density matrices, which are not included in either training or validation datasets. This dataset can be used directly as additional inputs for $\mathcal{L}_{\mathrm{phys}}$ in Eq.~\eqref{eq.phys_loss} during the training process and is re-generated at each epoch. Noticeably, since only the information at $t=0$ is needed, this approach expands the effective training set without solving any additional time evolution. Minimization of $\mathcal{L}_{\mathrm{phys}}$ for this on-the-fly sampled dataset enforces crucial algebraic properties of the neural propagator (i.e., its derivative should match ${\bm{L}}$ in Eq.~\eqref{eq.qme.mv} and its value at $t=0$ should give the identity), enabling it applicable to any initial state.

Finally, we present the validation test by focusing on the data loss $\mathcal{L}_{\mathrm{data}}$ only.
In Fig.~\ref{fig.validation}, we show the relative error of $\mathcal{L}_{\mathrm{data}}$ for each sample in the validation set. For all samples, the data loss is always around $\sim 1.0\%$.
To further reduce the validation error, a straightforward solution is to increase the size of the FNO propagator by using more layers and more hidden channels. 
We thus conclude that our constructed neural operator can be regarded as a universal propagator for any initial state and time $t$ within $[0, t_{\mathrm{max}}]$.

 
\section*{Data availability}
All the datasets in this work were generated directly from the code. They are also available from the corresponding authors upon reasonable request.

\section*{Code availability}
The code is available at \url{https://github.com/MrLightless/Neural-Quantum-Propagator}.

\section*{acknowledgement}
J.Z. and L.P.C. acknowledge support from the Key Research Project of Zhejiang Lab (No.~2021PE0AC02). C.L.B.-R. gratefully acknowledges the European Union’s Horizon Eu\-ro\-pe Re\-search and Innovation program  un\-der the Marie Skło\-dowska-Curie Grant Agreement n°101065295--RDMFTforbosons. Views and opinions expressed are however those of the authors only and do not necessarily reflect those of the European Union or the European Re\-search Executive Agency.

\bibliography{ref_new}

\end{document}